\documentclass[preprint,11pt]{elsarticle}
\usepackage{graphicx}
\usepackage[linesnumbered,boxed]{algorithm2e}
\usepackage{mathrsfs}
\usepackage{amssymb}
\usepackage{subfigure}
\input amssym.def

\begin{document}

\begin{frontmatter}
\title{Corner Occupying Theorem for the Two-dimensional Integral Rectangle Packing Problem}
\author{Wenqi Huang, Tao Ye \corref{cor1}}
\cortext[cor1]{Corresponding Author. Tel: 86-27-8754-3885;  Email: yeetao@gmail.com }
\address{School of Computer Science and Technology, Huazhong University of Science and Technology, Wuhan, 430074, China}
\author{Duanbing Chen}
\address{Web Sciences Center, School of Computer Science, University of Electronic Science and Technology of China, Chengdu, 611731, China}
\begin{abstract}
This paper proves a corner occupying theorem for the two-dimensional integral  rectangle packing problem, stating that if it is possible to orthogonally place $n$ arbitrarily given integral rectangles into an integral rectangular container without overlapping, then we can achieve a feasible packing by successively placing an integral rectangle onto a bottom-left corner in the container.  Based on this theorem, we might develop efficient heuristic algorithms for solving the integral rectangle packing problem. In fact, as a vague conjecture, this theorem has been implicitly mentioned with different appearances by many people for a long time.
\end{abstract}

\end{frontmatter}

\section{Introduction}
In the Integral Rectangle Packing (IRP) problem, we are given a set $J=\{1,2,\cdots,n\}$ of $n$ integral rectangles, each having width $ w_j$ and height $h_j$, and an integral rectangular container of width $W$ and height $H$. The problem is to determine whether all rectangles can be  orthogonally placed  into the container without overlapping. If the answer is yes, then we should present a non-overlapping packing pattern. In this paper, we assume that: (1) $w_j$, $h_j$, $W$ and $H$ are positive integers, and each vertex of each rectangle must be at integral point in the container. (2) Rectangles are rotatable, i.e., each rectangle can be horizontally or vertically placed into the container. 

The rectangle packing problem arises in many industrial applications, such as cutting wood, glass, paper and  steel in manufacturing, packing goods in transportation, arranging articles and advertisements in publishing. Various algorithms have been proposed to solve this problem. They can be divided into three categories: approximate algorithms, heuristic algorithms and exact algorithms \cite{hopper2001,lodi2002}. 

Most algorithms solve the RP problem by successively placing a new rectangle into the container. Then a basic problem arises:  where to place a new rectangle  when the container is already partially occupied by some previously placed rectangles?   To handle this problem, people have proposed several kinds of placement heuristics which specifies admissible positions for a rectangle \cite{baker1980, berkey1987, burke2004, huang2011, huang2007}. 

However, for a specific placement heuristic, there exists an important question: when there exist feasible solutions, can we achieve one by successively placing a rectangle into the container using this placement heuristic? If the answer is yes, then we say the placement heuristic is \textbf{complete}; otherwise, \textbf{incomplete}. If a placement heuristic is incomplete, then any algorithm based on it is foredoomed to fail on some instances. For example,  \citet{baker1980}  have proved that the Bottom-Left heuristic, which places a rectangle  onto the lowest possible position in the container and left-justify it, is incomplete. They found  an instance for which any feasible solution can not be achieved using the Bottom-Left heuristic, no matter what ordering of the rectangles is used.  \citet{martello2000} proposed a placement heuristic and developed an exact algorithm for the three dimensional bin packing problem. Later, \citet{boef2005} found that  some instances  can not be solved using the placement heuristic proposed by \citet{martello2000}.

It is usually very difficult to prove a placement heuristic's completeness or incompleteness. Nevertheless, for a special case of the RP problem, the 2D rectangular perfect packing problem, \citet{lesh2004} have shown that the  Bottom-Left heuristic is complete. They presented the following theorem: \textit{For every perfect packing, there is a permutation of the rectangles that yields that packing using the Bottom-Left heuristic.}  An efficient branch and bound algorithm is also developed based on this theorem. Besides this result, we have not found other paper in literature proving a certain placement heuristic's completeness. 
 
For the general IRP problem, this paper formulates a placement heuristic and proves its completeness. The following corner occupying theorem is presented that  \textit{ if it is possible to orthogonally place $n$ arbitrarily given integral rectangles into an integral rectangular container without overlapping, then we can achieve a feasible packing  by successively placing an integral rectangle onto a bottom-left corner in the container.} This theorem lays a solid foundation for understanding many efficient and exact algorithms for solving the RP problem\cite{christofides1977,hadji1995, kenmochi2009}.  It might be possible to develop new efficient and effective heuristic algorithms based on this theorem.

The rest of the paper is organized as follows. Section 2 presents several notations and definitions. Section 3 proves the corner occupying theorem.  Finally, Section 4 presents a counterexample to show why the proof of corner occupying theorem can not be directly extended to the three dimensional case. 
\section{Notations and definitions in integral rectangle packing problem}
We designate the bottom-left corner point of the container as the origin of the $xy$-plane and let its four sides parallel to $x$ and $y$ axis, respectively. The placement of rectangle $i (i=1,2,\cdots, n)$ in the container can be described by three variables $(x_i, y_i v_i)$, where $x_i, y_i \in \Bbb{N}=\{0,1,2,\cdots\} $ is the coordinate of its bottom-left corner point, $v_i  \in \{0,1\}$ denotes  its orientation. $v_i = 1$ means it is vertically placed, $v_i= 0$ horizontally. A packing pattern of $n$ rectangles can be described by a vector of $3n$ elements: $\mathcal{X}=(x_1, y_1, v_1, x_2, y_2, v_2,\cdots, x_n, y_n, v_n)$. We give the following definitions.

\newdefinition{mydef}{Definition}

\begin{mydef}[Feasible Packing]
A feasible (or non-overlapping) packing $\mathcal{X}$ satisfies the following two conditions:
\begin{enumerate}[(1)]
    \item Each rectangle does not overstep each border of the container.
    \item The overlapping area between any two rectangles is zero.
\end{enumerate}
\end{mydef}

\begin{figure}
\begin{center}
\includegraphics[width=1.2in]{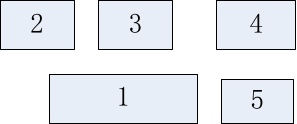}
\caption{Rectangles 2 and 3 are over rectangle 1}
\label{fig:j-over-i}
\end{center}
\end{figure}

\begin{mydef}[Rectangle $j$ Over Rectangle $i$]
We say rectangle $j$ is over rectangle $i$ (or rectangle $i$ is under rectangle $j$) if and only if there exists a positive number $d$ such that if rectangle $i$ moves upwards by a distance of $d$,  then the overlapping area between rectangles $i$ and $j$ is greater than zero.   See Fig.\ref{fig:j-over-i}, rectangles 2 and 3 are over rectangle 1, rectangles 4 and 5 are not over rectangle 1. We say rectangle $i$  \textit{ can move upwards freely} if and only if no rectangle is over rectangle $i$. 
\end{mydef}

\begin{figure}
\begin{center}
\includegraphics[height=0.9 in]{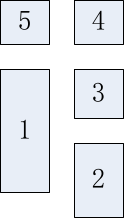}
\caption{Rectangles 2 and 3 are on the right of  rectangle 1}
\label{fig:j-on-the-right-of-i}
\end{center}
\end{figure}

\begin{mydef}[Rectangle $j$ On the Right of Rectangle $i$]
We say rectangle $j$  is on the right of rectangle $i$ (or rectangle $i$ is on the left of rectangle $j$) if and only if there exists a positive number $d$ such that if rectangle $i$ moves rightwards by a distance of $d$,  then the overlapping area between rectangles $i$ and $j$ is greater than zero. See Fig.\ref{fig:j-on-the-right-of-i}, rectangles 2 and 3 are on the right of rectangle 1, rectangles 4 and 5 are not on the right of rectangle 1.  We say rectangle $i$ \textit{can move rightwards freely} if and only if no rectangle is on the right of rectangle $i$. 
\end{mydef}

\begin{figure}
\begin{center}
\includegraphics[width=3in]{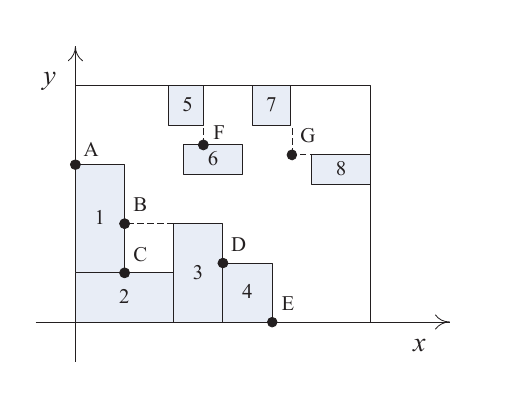}
\caption{Bottom-left stability and bottom-left corners}
\label{fig:blc}
\end{center}
\end{figure}

\begin{mydef}[Bottom-Left Stability]
In a feasible packing, a rectangle is \textit{ bottom-left stable} if and only if it can not move downwards or leftwards without overlapping others \cite{chazelle1983}. A feasible packing is \textit{bottom-left stable} if and only if each rectangle in this packing is bottom-left stable. See Fig.\ref{fig:blc},  rectangles 1, 2, 3, 4 are bottom-left stable; rectangles 5, 6, 7, 8 are not bottom-left stable.
\end{mydef}

\begin{mydef}[Bottom-Left Corner]
 A bottom-left corner is an unoccupied area where a suitable rectangle has bottom-left stability. See Fig.\ref{fig:blc}, A, B, C, D, E, F, G are bottom-left corners.
\end{mydef}

\begin{mydef}[Corner Occupying Action]
A corner occupying action is an action that places a rectangle onto a bottom-left corner and makes that rectangle bottom-left stable.  Let the placed rectangle be rectangle $i$, then \textit{the rectangles forming the bottom-left corner where rectangle $i$ locates} satisfy the following two conditions: (1) they touch rectangle $i$; (2) they are under or on the left of rectangle $i$.   
\end{mydef}

\section{Corner Occupying Theorem}
This section proves the corner occupying theorem. We first present two lemmas.

\newtheorem{mylem}{Lemma}
\newproof{myprf}{Proof}

\begin{mylem}
Any feasible packing can be replaced by another feasible packing where each rectangle has bottom-left stability.
\end{mylem}

\begin{figure}
\begin{center}
 \subfigure[Not bottom-left stable]{
        \includegraphics[width=1.4in]{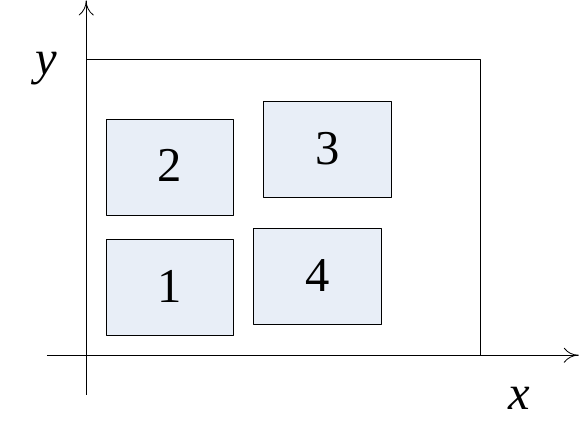}}
  \subfigure[Bottom-left stable]{
        \includegraphics[width=1.4in]{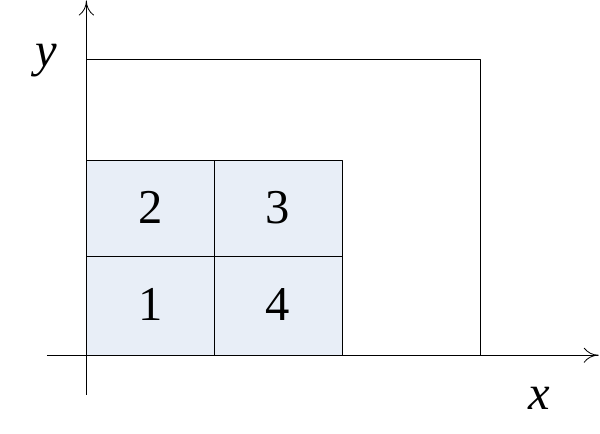}}
  \end{center}
  \caption{A feasible packing and its equivalent bottom-left stable packing}
  \label{fig:lemma1}
\end{figure}

\begin{myprf}
See Fig.\ref{fig:lemma1}, the left packing can be replaced by the right bottom-left stable one. Given a feasible packing $\mathcal{X}_0$, we prove that an equivalent bottom-left stable packing can be found from $\mathcal{X}_0$. Keep the orientation of each rectangle unchanged, suppose that each rectangle can move freely and consider the following function:
\begin{equation}
O = O(x_1, y_1, x_2, y_2, \cdots, x_n, y_n) = \sum_{i=0}^{n-1}{ \sum_{j=i+1}^{n}{O_{ij}} }
\end{equation}
where $O_{ij} (i,j=1,2,\cdots,n)$ is the  overlapping area between rectangles $i$ and $j$. $O_{0j} (j=1,2,\cdots, n)$  is the overlapping area between rectangle $j$ and the \textbf{outside} of the container.  Let $S_0$ be the set of all zero points of $O$: $S_0 = \{ (x_1, y_1, x_2, y_2, \cdots, x_n, y_n)   |  O(x_1, y_1, x_2, y_2, \cdots, x_n, y_n) = 0 \}$. Then each point in $S_0$ corresponds to a non-overlapping packing.  Because $\mathcal{X}_0$ corresponds to a zero point of $O$,  $S_0$ is not empty. And because each rectangle can only be placed at integer coordinate positions, $S_0$ is a finite set. 

Then let's consider another function $L$ defined on $S_0$:
\begin{equation}
   L = \sum_{i=1}^{N} (x_i + y_i)
\end{equation}
Because $S_0$ is a non-empty and finite set, there exists a point  $(x_1^*, y_1^*, x_2^*, y_2^*, \cdots, x_n^*, y_n^*)$ in $S_0$ where $L$ is minimal.  Note that $(x_1^*, y_1^*, x_2^*, y_2^*, \cdots, x_n^*, y_n^*)$  corresponds to a feasible packing where each rectangle can not move downwards or leftwards without overlapping others; otherwise, we can find another point in $S_0$ with a smaller $L$, contradicting  the fact that $L$ attains its minimum at  $(x_1^*, y_1^*, x_2^*, y_2^*, \cdots, x_n^*, y_n^*)$.  $\square$
\end{myprf}

Lemma 1 has been explicitly mentioned by \cite{christofides1977, hadji1995, martello2000}  as a conjecture and implicitly used by many algorithms for solving the rectangle packing problem. 

\begin{mylem}[Escaping Lemma]
In any feasible packing, if we take away the four borders of the container, then there is a rectangle which can move upwards and rightwards freely.
\end{mylem}

\begin{figure}
\begin{center}
  \subfigure[]{
      \includegraphics[width=1.4in]{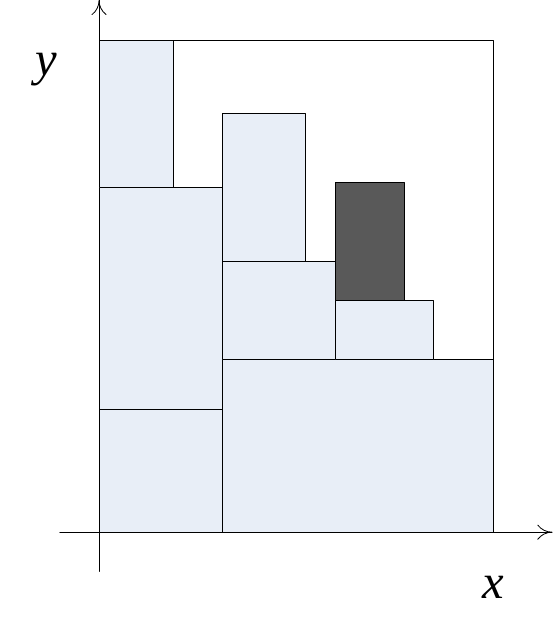}}
  \subfigure[]{
      \includegraphics[width= 1.4in]{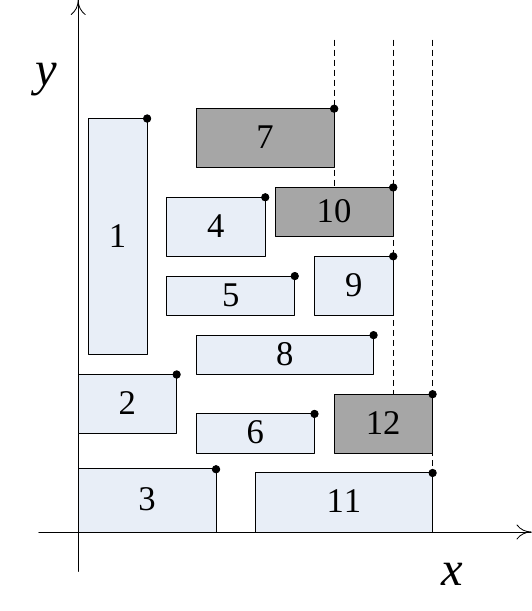}}
  \caption{Examples for Lemma 2}
  \label{fig:lemma2}
\end{center}
\end{figure}
 \begin{myprf}
See Fig.\ref{fig:lemma2}(a), the highlighted rectangle can move  upwards and rightwards freely. Given a feasible packing with $n$ rectangles, we sort the  top-right corner points of the rectangles lexicographically by increasing $<x,y>$ and renumber the rectangles according to this order (See Fig.\ref{fig:lemma2}(b)). We search for the rectangle which can move rightwards and upwards freely as follows. First, we consider the highest numbered rectangle among all $n$ rectangles, i.e., rectangle $n$ (12 in Fig.\ref{fig:lemma2}(b)). Its top-right corner point is the rightmost, thus it can move rightwards freely. If no rectangle is over rectangle $n$,  then $n$ is the rectangle we want to find.  Otherwise,  we consider the highest numbered rectangle among all the rectangles over rectangle $n$. Let it be rectangle $i$ (10 in Fig.\ref{fig:lemma2}(b)). Its top-right corner point is the rightmost among all the rectangles over rectangle $n$. Therefore, it can move rightwards freely. If no rectangle is over rectangle $i$, then $i$ is the rectangle we want to find. Otherwise we consider the highest numbered rectangle among all the rectangles over rectangle $i$ and continue the search as described above.
 
Because there are only finite ($n$) rectangles, the above search will terminate  and we can finally find a rectangle which can move upwards and rightwards freely.  $\square$
\end{myprf}

\newtheorem{mythem}{Theorem}
\begin{mythem}
For any feasible, bottom-left stable packing, there exists a numbering of $n$ rectangles such that rectangle $i(i=1,2,\cdots,n)$ locates on a bottom-left corner formed by rectangles $1,2,\cdots, i-1$ and the four borders of the container.
\end{mythem}

\begin{figure}
\begin{center}
  \subfigure[Numbering According to Lemma 2]{
      \includegraphics[width=1.4in]{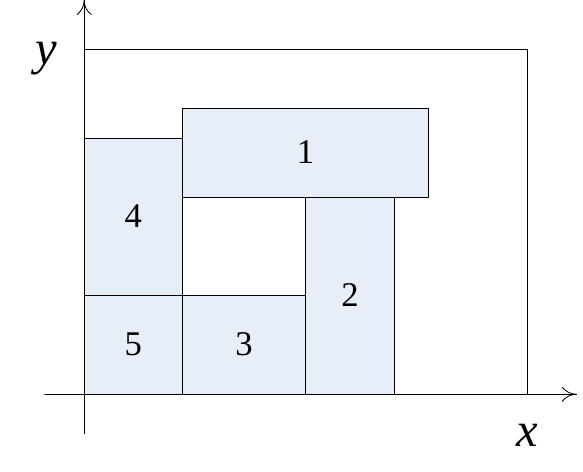}}
  \subfigure[New Numbering]{
      \includegraphics[width= 1.4in]{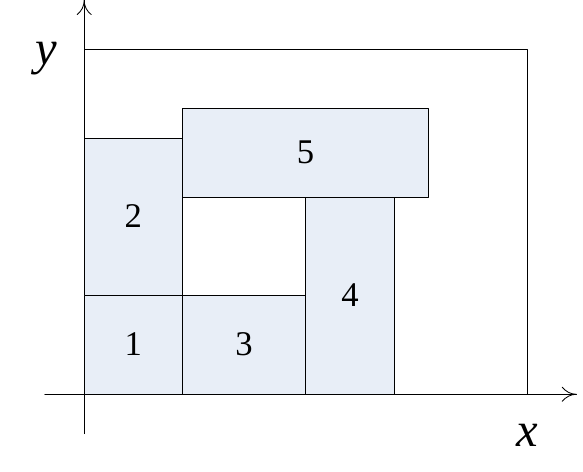}}
  \caption{Examples for Theorem 1}
  \label{fig:theorem1}
\end{center}
\end{figure}

\begin{myprf}
According to Lemma 2, there is always a rectangle which can move rightwards and upwards freely in any feasible packing. Consequently, for a feasible  packing with $n$ rectangles,  we can empty the container by successively taking out a rectangle which can move rightwards and upwards freely. We then number the rectangles according to the order in which rectangles are taken out. Under this numbering, a higher numbered rectangle is not  over or on the right of a lower numbered rectangle (See Fig.\ref{fig:theorem1}(a)).

Now we consider a new numbering which is the reversion of the previous numbering, i.e., the number of rectangle $i(i=1,2,\cdots,n)$ becomes $n-i+1$ (See Fig.\ref{fig:theorem1}(b)). Under this new numbering, a lower numbered rectangle is not over or on the right of a higher numbered rectangle. Then, for  rectangle $i(i=1,2,\cdots,n)$, the labeled numbers of the rectangles forming the bottom-left corner where rectangle $i$ locates are smaller than $i$.

The new numbering of the rectangles can be taken as an order in which rectangles are placed into the container. Under this order, when we place rectangle $i$ into the container, rectangles $1,2,\cdots, i-1$ are already in the container and rectangles $i+1, i+2, \cdots, n$ are not.  Thus, it is shown that, any feasible, bottom-left stable packing can be achieved through a sequence of placement actions, among which the $i$th $(i=1,2,\cdots, n)$ action is to place  rectangle $i$ onto a bottom-left corner formed by rectangles $1,2,\cdots, i-1$ and the four borders of the container. That is to say, any feasible, bottom-left stable packing can be achieved through a sequence of corner occupying actions.
$\square$
\end{myprf}

\begin{mythem}[Corner Occupying Theorem]
Arbitrarily given $N$ rectangles and a rectangular container, if it is possible to orthogonally place all the rectangles into the container without overlapping, then we can find a feasible packing through a sequence of corner occupying actions.
\end{mythem}
\begin{myprf}
According to Lemma 1, there exists a feasible packing where each rectangle has bottom-left stability. Then according to Theorem 1, this bottom-left stable packing can be found through a sequence of corner occupying actions. $\square$
\end{myprf}

Note that, in the above proof, the escaping lemma determines the order in which rectangles are placed into the container. And we can find that, when we place rectangle $i$ onto a bottom-left corner in the container, no rectangle in the container is  over or on the right of rectangle $i$.  Therefore, in practical implementation, in order to reduce the computing time,  the corner occupying theorem can be enhanced as : \textit{Arbitrarily given $N$ rectangles and a rectangular container, if it is possible to orthogonally place all the rectangles into the container without overlapping, then we can find a feasible packing by successively placing a rectangle onto a bottom-left corner in the container where  the placed rectangle is then under and on the left of no rectangle .}  

\section{The Three-dimensional Case}
In this section, we investigate the following problem: can the lemmas and theorems presented in Section 3 be extended to the three-dimensional case?  Scientists have done some research for the three-dimensional case\cite{boef2005,martello2000}.

In the three-dimensional case, rectangle $j$ has width $w_j$, height $h_j$ and depth $d_j$, and the container is of width $W$, height $H$ and depth $D$.  Similarly, we can extend the notations and definitions presented in Section 2 to the three-dimensional case, and get new definitions like: rectangle $j$ in front of rectangle $i$ (or rectangle $i$ in behind of rectangle $j$), bottom-left-behind corner,  bottom-left-behind stability. And then the corner occupying action is to place a rectangle onto a bottom-left-behind corner in the container.

\begin{figure}
\begin{center}
\includegraphics[width=3in]{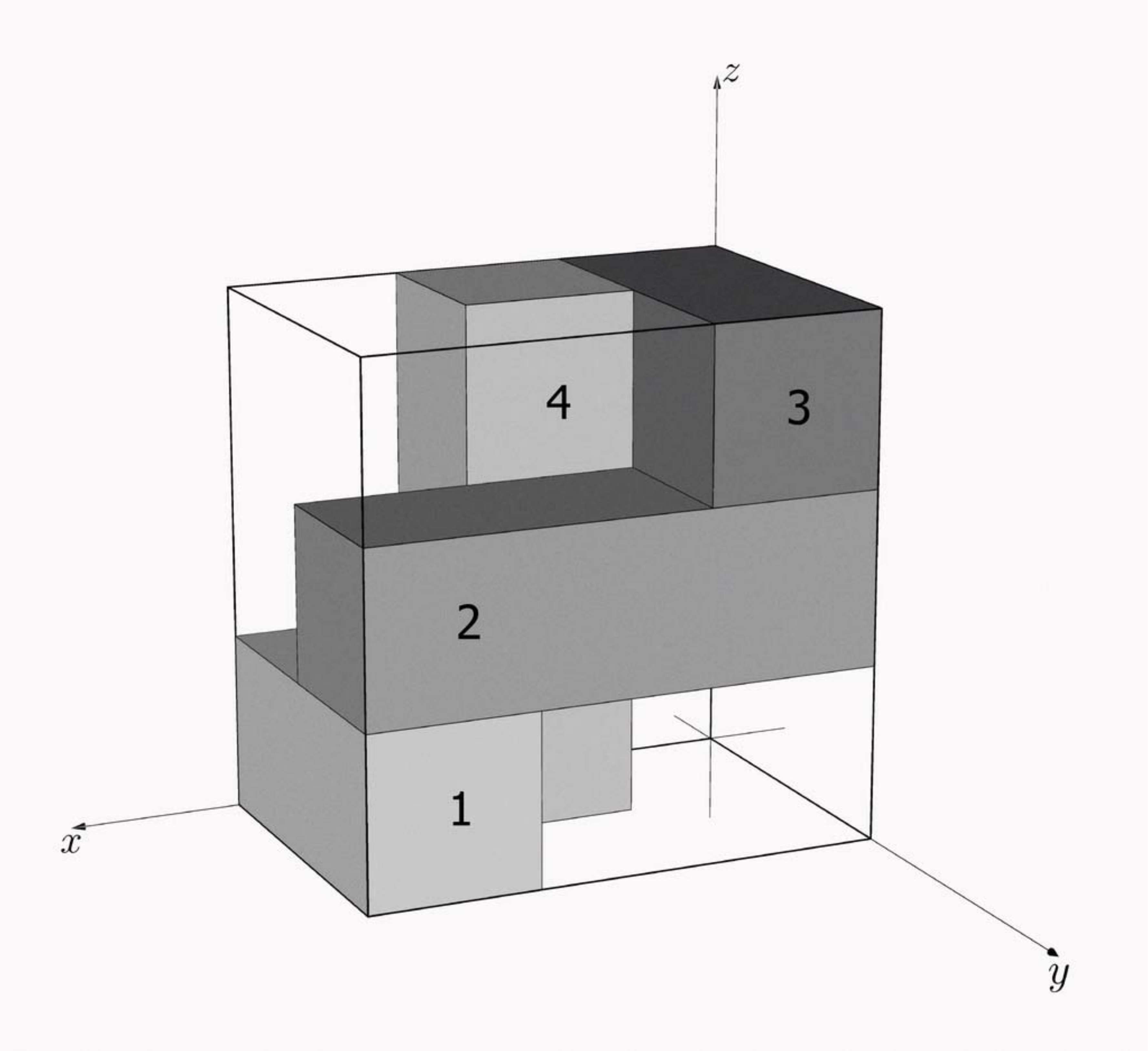}
\caption{A counterexample of escaping lemma in the three-dimensional case}
\label{fig:3d}
\end{center}
\end{figure}

We find that lemma 1 and its proof can be easily extended to the three-dimensional case. However, the escaping lemma is wrong in the three-dimensional case. Fig.\ref{fig:3d} presents a counterexample where no rectangle can move along the positive x-axies, positive y-axies, and positive z-axies freely. The coordinate position of each rectangle in Fig.\ref{fig:3d} is presented in Table \ref{tbl:3d}. 
\begin{table*}
\caption{Coordinate position of each rectangle in Fig.\ref{fig:3d} }
\label{tbl:3d}
\begin{center}
\begin{tabular}[htbp]{ccccccc} 
\hline 
  Rectangle &  $x$ &  $y$ & $z$ & $width$ & $depth$ & $height$ \\ 
\hline 
  1 & 2 & 0 & 0 & 1 & 2 & 1 \\  
  2 & 0 & 1 & 1 & 3 & 1 & 1 \\  
  3 & 0 & 0 & 2 & 1 & 2 & 1 \\  
  4 & 1 & 0 & 0 & 1 & 1 & 3 \\  
\hline
\end{tabular}
\end{center}
\end{table*}

As shown in Section 3,  in the two-dimensional case, the proof of the corner occupying theorem is based on lemma 1 and the escaping lemma. Since the escaping lemma is wrong in the three-dimensional case, we can not directly extend the proof of the corner occupying theorem to the three-dimensional case. However, the incorrectness of escaping lemma in the three-dimensional case might not imply that the corner occupying theorem is incorrect in the three-dimensional case. Therefore, we get an open problem: is the corner occupying theorem correct in the three-dimensional case? Or specifically,

\textit{In the three-dimensional case,  if it is possible to orthogonally place $n$ arbitrarily given rectangles into a rectangular container without overlapping, can we achieve a feasible packing by successively placing a rectangle onto a bottom-left-behind corner in the container?}

\end{document}